\documentclass{aa}  
\usepackage{natbib} 
\bibpunct{(}{)}{;}{a}{}{,} % to follow the A&A style
\usepackage{graphicx}
\usepackage{txfonts}
\usepackage[colorlinks, linkcolor=blue, citecolor=blue, urlcolor=blue]{hyperref} 
\usepackage{placeins}
\begin{document}

   \title{A possible GeV-TeV connection in the direction of the Globular Cluster UKS 1}

   \author{Jiwon Shin,
          \inst{1}
          C. Y. Hui,\inst{2}\fnmsep\thanks{Corresponding author: E-mail: cyhui@cnu.ac.kr, huichungyue@gmail.com}
          Sangin Kim,\inst{1}
          Kwangmin Oh,\inst{3}
          \and
          Ellis R. Owen\inst{4, 5}
          }

   \institute{Department of Earth, Environmental \& Space Sciences, Chungnam National University, Daejeon 34134, Republic of Korea
   \and
   Department of Astronomy and Space Science, Chungnam National University, Daejeon 34134, Republic of Korea
    \and
    Department of Physics and Astronomy, Michigan State University, East Lansing, MI 48824, USA
    \and
    Astrophysical Big Bang Laboratory (ABBL), RIKEN Cluster for Pioneering Research, Wak\={o}, Saitama, 351-0198, Japan
    \and 
    Theoretical Astrophysics, Department of Earth and Space Science, Graduate School of Science, Osaka University, Toyonaka, Osaka 560-0043, Japan}

  \abstract
  % context heading (optional)
  % {} leave it empty if necessary  
   {}
  % aims heading (mandatory)
   {Using public data collected by the {\it Fermi} Large Area Telescope (LAT) over 16 years, and the 1523 days of survey data (3HWC) from the High Altitude Water Cherenkov (HAWC) observatory, we searched for possible GeV-TeV connections in globular clusters (GCs).}
  % methods heading (mandatory)
   {Excluding 44 confirmed $\gamma-$ray detections of GCs in the latest Fourth {\it Fermi}-LAT point source catalog (4FGL-DR4), we  searched for possible GeV emission from the other 113 known GCs using 16 years of LAT data. We performed a systematic binned likelihood analysis in the energy range 0.3-100 GeV towards these targets. We also searched for possible TeV excesses in the directions of 27 GeV-detected GCs covered by the 3HWC survey area, assuming a point-source morphology and a power-law spectrum of $E^{-2.5}$.  }
  % results heading (mandatory)
   {In addition to the confirmed $\gamma-$ray GCs in the 4FGL catalog, we report a GeV detection at the position of UKS 1 with a post-trial probability of $\sim8\times10^{-5}$ of it being a fluctuation. Its spectrum within this energy range is well described by a power-law model with $\Gamma\simeq2.3\pm0.5$. Furthermore, this GeV feature appears to extend southeast in a direction towards the Galactic plane. From the 3HWC survey data, we have also identified a TeV feature in the direction of UKS 1. It is well-resolved from any known Very High Energy (VHE) source. The post-trial probability that this feature is a fluctuation is $\sim3\times10^{-4}$. If confirmed, this would be the second detection of a TeV feature in the proximity of a GC. While the GeV emission mostly coincides with the center of UKS 1, the TeV peak is displaced from the cluster center by several tidal radii in the trailing direction of the GC's proper motion. Given the supersonic speed of UKS 1 at $\sim270$~km~s$^{-1}$, our findings are consistent with a scenario where the VHE $\gamma-$rays are produced by inverse Compton scattering between relativistic particles and ambient soft photon fields during the course of their propagation away from the head of the bow shock.}
  % conclusions heading (optional), leave it empty if necessary 
   {}

\titlerunning{Searches for GCs in the GeV-TeV regime} 
  
   \maketitle
%
%-------------------------------------------------------------------

\section{Introduction}
Shortly after the {\it Fermi} Large Area Telescope (LAT) commenced operations, $\gamma-$ray emission from the globular cluster (GC) 47 Tuc was detected \citep{47Tuc_LAT}. Subsequently, a number of other $\gamma-$ray GCs were soon revealed, establishing them as a unique source class in the $\gamma-$ray sky \citep[e.g.][]{GC_LAT,Tam_2011}. This includes Terzan 5, which hosts the largest known population of millisecond pulsars (MSPs) \citep{Kong_2010}, which are considered to be the origin of the $\gamma-$ray emission. 

The $\gamma-$ray spectra of most GCs resemble those of MSPs, which can be phenomenologically characterized by a power-law model with an exponential cutoff (PLEC) \citep[e.g.][]{GC_LAT}. Except for a few cases where individual luminous MSPs dominate the emission from a GC  \citep{NGC6624A,Wu_2013,NGC6652B,M92A}, the $\gamma-$rays from GCs are likely to be a collective contribution from the whole MSP population residing within the cluster. 
This is supported by reported correlations between $\gamma-$ray luminosity of GCs, $L_{\gamma}$, and the primary stellar encounter rate within their cores, $\Gamma_{\rm GC}$, which resembles the correlation between the MSP population and $\Gamma_{\rm GC}$ \citep[e.g.][]{Hui2010.714, Hui_2011,GC_LAT}.  However, a recent study does not find a significant correlation between $L_{\gamma}$ and $\Gamma_{\rm GC}$ \citep{2024stackGC} which indicates the relation between MSPs and $L_{\gamma}$ may be more complex.

$\gamma-$rays can also be produced through inverse Compton scattering (ICS) between relativistic leptons in the pulsar wind and ambient soft photon fields \citep[e.g.][]{Cheng_2010}. The ICS scenario is supported by a possible correlation between GC $L_{\gamma}$ and their in-situ soft photon energy densities \citep{Hui_2011, 2024stackGC, Song_2021}, as well as evidence of a power-law tail (PL) in addition to the PLEC component of their $\gamma-$ray spectra \citep{Song_2021}. 

ICS between pulsar wind and soft photons is also expected to produce Very High Energy (VHE $>0.1$~TeV) photons \citep[e.g.][]{Cheng_2010}. However, the TeV feature found in the direction of Terzan 5 remains the only confirmed detection of such emission \citep{Ter5_HESS}. Moreover, an offset between the TeV emission peak and the center of the cluster has cast doubt on its physical association.

Theoretical studies have attempted to explain how a GC can produce misaligned TeV features. \cite{Bednarek_2014} proposed a scenario where a GC with a characteristic space velocity of order $\sim100$~km~s$^{-1}$ forms a bow shock nebula around a GC where the mixture of stellar and pulsar winds leaving a GC interacts with the surrounding interstellar medium. For a GC located close to the Galactic plane, like Terzan 5, a nebula with asymmetric morphology is expected from the formation of a bow shock ahead of the GC's motion in the dense surrounding medium. Particle advection in the relativistic wind opposite to the direction of the GC's motion then produces a  misalignment between the TeV feature and the GC center. 

This does not explain why the peak of the TeV emission does not coincide with the acceleration site of the particles driving the ICS emission. The peak energy density of the accelerated particles should be located at the wind termination shock. There, the starlight from the GC produces a strong ambient photon field capable of forming an ICS TeV peak. However, the relativistic particles emanating from the bow shock are expected to be strongly oriented along the trailing magnetotail, with anisotropic ICS emission directed along their propagation vector. Self-scattering leads to a gradual broadening of the particle pitch angle distribution, isotropizing their emission and allowing it to become visible some distance away from the acceleration site \citep{Krumholz_2024}. In this sense, such systems can be used to constrain the scattering processes of cosmic rays in the Milky Way.   

In this study, we aim to search for a possible GeV-TeV connection in the directions of GCs in Milky Way by using the data obtained by {\it Fermi}-LAT, and from a high sensitivity survey of the northern VHE sky conducted by the High Altitude Water Cherenkov (HAWC) observatory.

 \begin{figure*}
  \centering
\includegraphics[width=18cm]{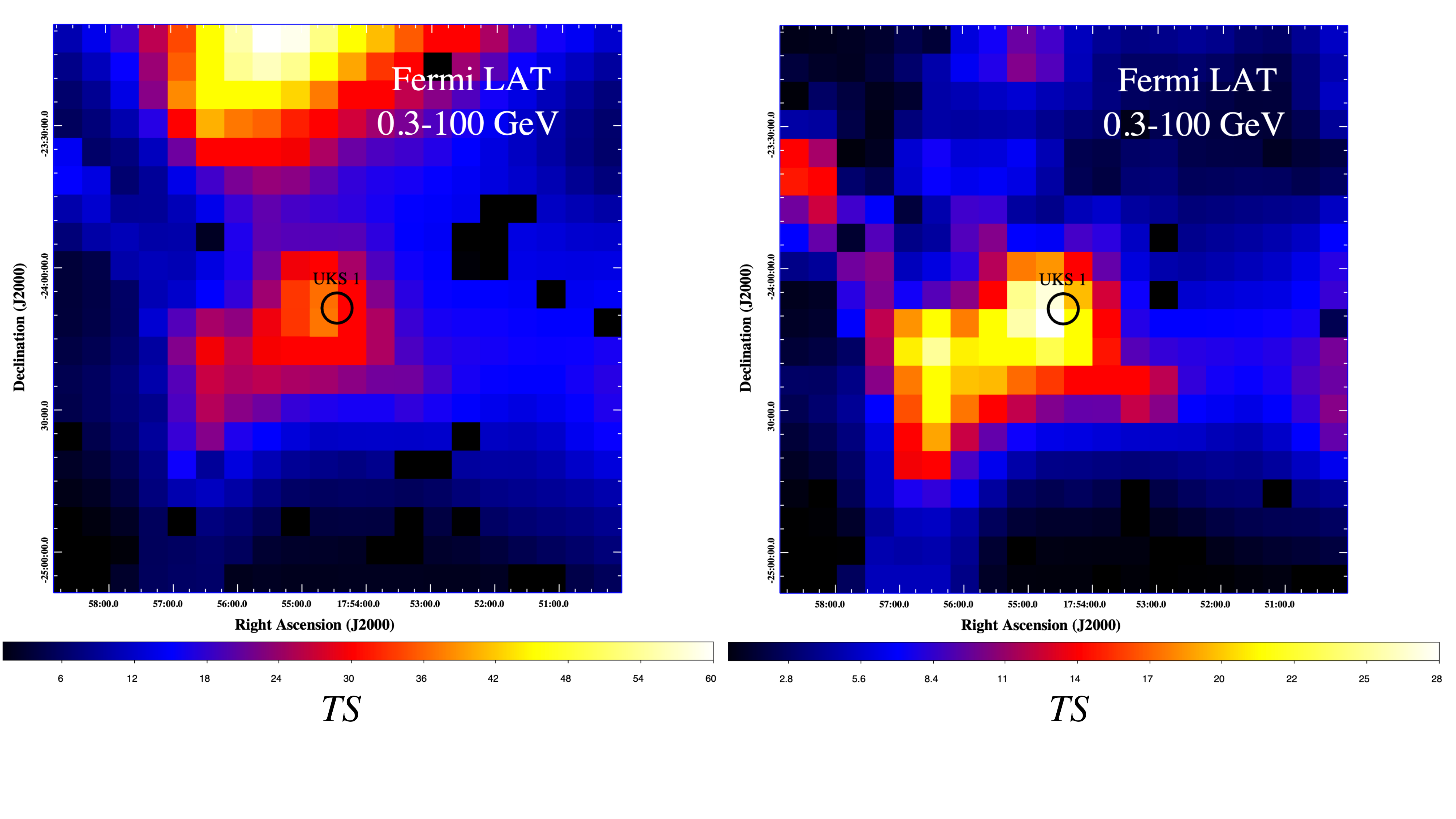}
\vspace{-1cm}
   \caption{{\it Left panel}: TS map of the $2^{\circ}\times2^{\circ}$ region centered on UKS 1 with the best-fit diffuse background and contributions from the 4FGL sources subtracted. The black circle represents the tidal radius ($3.2^{'}$) of  UKS 1 at a distance of 15.6~kpc \citep{uks1_pm}. The colour scale shows the TS value of every bin of $0.1^{\circ}\times0.1^{\circ}$. {\it Right panel}: Same as left panel but with three additional sources further subtracted  (see Appendix~\ref{excess}).}
   \label{lat_ts}
   \end{figure*}
%-----------------------------------------------------------------
%----------------------------------------------------------------- 
 \begin{figure}
  \centering
\includegraphics[width=9cm]{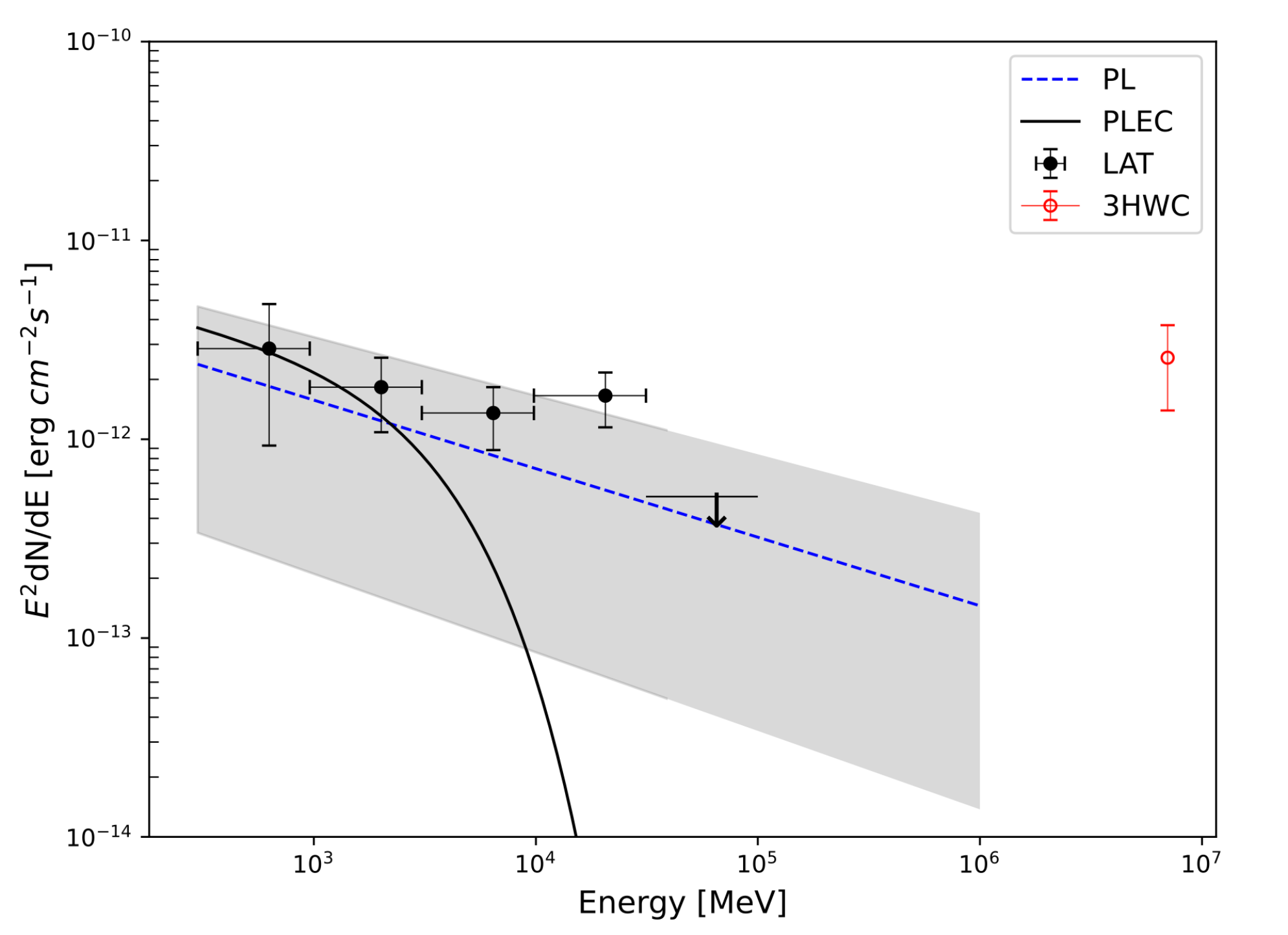}
   \caption{Spectral energy distribution (SED) of the $\gamma-$ray emission originating from the direction of UKS 1, as observed by {\it Fermi}-LAT (black data points). The blue dashed-line is the best-fit PL model. The shaded region illustrates the uncertainty of this model. The PLEC fit is also plotted for comparison. Both models and LAT data in this plot are the results with the surrounding excess sources removed. The estimated energy flux derived from the differential flux given by the 3HWC data at 7 TeV is shown in red.}
   \label{sed}
   \end{figure}
%-----------------------------------------------------------------
%----------------------------------------------------------------- 
 \begin{figure}
  \hspace{-0.8cm}
\includegraphics[width=10cm]{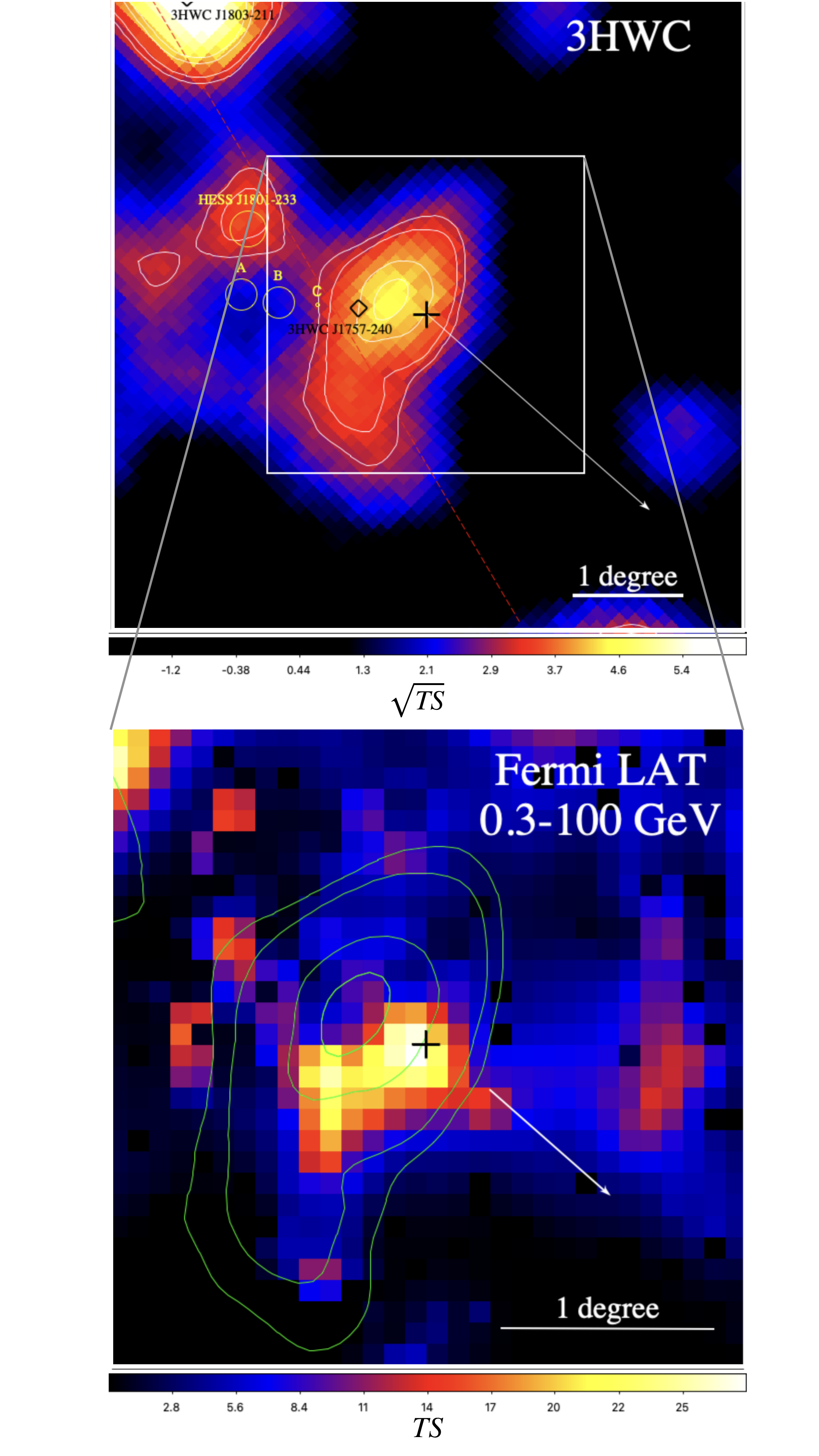}
   \caption{{\it Top panel:} $6^{\circ}\times6^{\circ}$ 3HWC significance map around the center of UKS 1 from a point-source search with contours illustrating significance levels of $3\sigma$, $3.3\sigma$, $4\sigma$ and $4.3\sigma$. The location and extent of nearby TeV sources are indicated by yellow circles. The red dashed lines shows the orientation of Galactic plane. {\it Bottom panel:} A close-up view within $3^{\circ}\times3^{\circ}$ of the {\it Fermi}-LAT TS map, with 3HWC contours overlaid.  The black crosses and white arrows in both panels illustrate the center and the direction of the proper motion of UKS 1 respectively. Top is north and left is east. A scale bar of $1^{\circ}$ is given in both panels.}
   \label{hwc_lat}
   \vspace{-0.3cm}
   \end{figure}

\section{Data Analysis and results}
\subsection{Searches for additional GeV-detected GC with {\it Fermi}-LAT data}
Excluding the 44 confirmed GeV-detected GCs in the most updated Fourth {\it Fermi}-LAT point source catalog \citep[][4FGL-DR4]{4fgldr4}, i.e. 41 with positional coincidence plus 3 with confirmed $\gamma-$ray pulsations, we have searched for possible $\gamma-$rays from the other 113 listed by \citet[][2010 edition]{Harris1996.112}. 

In this work, we used {\it Fermi}-LAT data collected from 2008-08-04  to 2024-06-23, covering $\sim16$~years. The {\it Fermitools}-2.2.0 package\footnote{\url{https://github.com/fermi-lat/Fermitools-conda/}} was used for data reduction and analysis. To search for $\gamma-$rays from the target GCs, we selected the events from a region-of-interest (ROI) within a 10$^{\circ}$ radius from of the GC locations given by \citet[][2010 edition]{Harris1996.112}. Following the standard {\it Fermi}-LAT data analysis thread, only the events classified as ``Source" class (evclass = 128) and event type ``FRONT+BACK" (evtype =3) were included. Since photons coming from the Earth limb can produce a strong background, we minimize their possible contamination by applying a zenith angle cut of $<90^{\circ}$.  To further assess the data quality and to select good time intervals (GTIs), we filter the data with the expressions (DATA\_QUAL > 0) \&\& (LAT\_CONFIG == 1). Throughout this study, the instrument response function (IRF) P8R3\_SOURCE\_V3 was used. All the analyses were performed in the energy range of 0.3-100 GeV. 

Using the {\it Fermitools} task {\it gtlike}, we performed a binned maximum likelihood analysis for the data around each target GC. To model all the point and extended sources in 4FGL-DR4 catalog within the corresponding ROI of the target, we adopted their spectral parameters given in the catalog as initial values. For 4FGL sources within $5^{\circ}$ from the center of the ROI, we thawed their spectral parameters for fitting. For the other 4FGL sources, their parameters were fixed in the analysis. In view of accounting for the wide point spread function of LAT, particularly at low energies, we also included the 4FGL sources in an extra $10^{\circ}$ annular region outside the ROI in the likelihood analysis. Apart from the resolved 4FGL sources, we have also modeled the Galactic diffuse background (gll\_iem\_v07.fits) and the extragalactic isotropic background (iso\_P8R3\_SOURCE\_V3\_v1.txt)  \citep{gamma_bkg}.
For all the targeted GCs, we assume that they are point-sources with a PL spectrum. 

We set the detection threshold at $4\sigma$ and require the target to be well resolved from its surroundings. Among the 116 GCs we have analyzed, only UKS 1 satisfied our detection criteria. With the aforementioned model, our likelihood analysis yields a test statistic (TS) of 37 \citep[see Sec. 3 in][for the definition of TS]{1996ApJ...461..396M}. Taking the number of GCs we have searched as the trial factor (i.e. 113), the observed TS value corresponds a post-trial probability $\sim10^{-6}$ that the feature is a background fluctuation. 
The best-fit PL model has a photon index of $\Gamma=2.50\pm0.50$. In contrast, a PLEC fit yields a lower TS value of 29.3. While the photon index of the best-fit PLEC model is comparable to that of the PL model ($\Gamma=2.24\pm0.21$), the cut-off energy $E_{c}=3.0\pm6.5$~GeV cannot be constrained properly. These results suggest that the PLEC model cannot provide an adequate description of the data. 

To examine the surroundings, we have computed the TS map centered on UKS 1 with the {\it Fermitools} task {\it gttsmap}. A $2^{\circ}\times2^{\circ}$ TS map is shown in the left panel of Figure~\ref{lat_ts}. A $\gamma-$ray excess at the location of the GC is clear. This confirms the findings first reported by \cite{Yuan_2022} at a significance of $\sim4\sigma$. Examination of the TS map suggests that the $\gamma-$rays associated with UKS 1 appears to be extended towards the southeast by $\sim0.5^{\circ}$ with a somewhat lower TS value. 
 
Besides UKS 1, we have also identified several $\gamma-$ray excesses with the ROI (see Appendix~\ref{excess}).
With these excesses further subtracted, the TS value at the center of UKS 1 goes down to 28.4 for the PL model, which corresponds to a post-trial probability $7.7\times10^{-5}$ as a fluctuation. 
In the right panel of Figure~\ref{lat_ts}, we show the TS map with the aforementioned excesses subtracted. It clearly shows the feature associated with UKS 1 extends towards southeast. Its spectral energy distribution (SED) is shown in Figure~\ref{sed}. The blue dashed-line represents the best-fit PL spectrum with a 
photon index of $\Gamma=2.34\pm0.50$. Integrating the model between 0.3-100 GeV yields an energy flux of $f^{\rm LAT}_{\gamma}=(3.24\pm2.74)\times10^{-11}$~erg~cm$^{-2}$~s$^{-1}$. At the distance of UKS 1, 
$d=$~15.6~kpc \citep{Baumgardt2021MNRAS},  
this translates to a $\gamma-$ray luminosity of 
$L^{\rm LAT}_{\gamma}=(9.44\pm7.97)\times10^{35}$~erg~s$^{-1}$. 
We also examined the PLEC fit after excess subtraction. The model parameters remained largely unchanged compared to the pre-subtraction fit, except that the TS value decreased further to 18.3. The cutoff energy also remains unconstrained. In light of these results, we do not consider the PLEC model further in this work.
 
\subsection{Search for TeV feature in the direction fo GeV-detected GCs with 3HWC Survey data}

To search for possible TeV features associated with GeV-detected GCs, we used data collected by HAWC during the survey with its complete instrumental configuration on the first 1523 days of operation.\footnote{\url{https://data.hawc-observatory.org/datasets/3hwc-survey/fitsmaps.php}} This survey led to the compilation of the 3$^{\rm rd}$ HAWC catalog of VHE $\gamma-$ray sources \citep[3HWC;][]{3hwc}. This reported 65 sources detected at a significance $\geq5\sigma$, being the most sensitive TeV survey of the northern sky.  

We started the search by cross-correlating the sources in the 3HWC catalog with the positions of GeV-detected GCs. The location of HAWC allows it to cover the sky within declinations from around $-26^{\circ}$ to $64^{\circ}$. Of the 45 GeV-detected GCs (44 in 4FGL-DR4, plus UKS 1), only 27 fall within the survey area. In an extended-source search assuming a disk-like morphology with a radius of $1^{\circ}$ \citep{3hwc}, the source 3HWC~J1757-240 in the 3HWC catalog is located $\sim38^{'}$ east from the center of UKS 1. The extension of this source overlaps with the complex W28 field that encompasses four known TeV sources: HESS~J1801-233 and HESS~J1800-240A/B/C \citep{hess}. 

While the interactions between the supernova remnant W28 and surrounding molecular clouds can produce $\gamma-$rays \citep[e.g.][]{w28}, we speculate the low angular resolution of the extended-source search as reported by \cite{3hwc} might blend several sources into a single extended source, leading to a bias in identifying the source nature. To address this issue, we performed a search using the 3HWC significance map resulting from a point-source search, assuming an power-law energy spectrum with an index of 2.5. In the top panel of Figure~\ref{hwc_lat}, we show a $6^{\circ}\times6^{\circ}$ 3HWC significance map centered on UKS 1. To better illustrate the significance variation, we have drawn the contours for significance levels of $3\sigma$, $3.3\sigma$, $4\sigma$ and $4.3\sigma$. The peak significance of the feature in proximity to UKS 1 is located at 
RA=$17^{\rm h}55^{\rm m}54.38^{\rm s}$, Dec=$-23^{\circ}57^{'}30.96^{''}$ (J2000), and has a significance of $4.4\sigma$. 
With a trial factor of 27, this corresponds to a post-trial probability of $2.9\times10^{-4}$ as fluctuation. 
It is displaced from the center of UKS 1 by $\sim23^{'}$ in the direction of N60.7$^{\circ}$E. Interestingly, we found that the displacement almost traces the direction opposite to the proper motion of UKS 1 \citep[$(\mu_{\alpha}\cos\delta,\mu_{\delta})=(-2.77,-2.43)$~mas~yr$^{-1}$;][]{uks1_pm}, i.e. towards S48.7$^{\circ}$W. The direction of the GC's proper motion is illustrated by the arrow in Figure~\ref{hwc_lat}. 

The location and extension of the nearby TeV sources found by \cite{hess} are indicated by yellow circles in the top panel of Figure~\ref{hwc_lat}. This shows that the feature associated with UKS 1 is well-resolved from nearby TeV sources. HESS~J1801-233 is identified as a separate source in this significance map resulting from a point-source search, and there is no evidence for any excess at the locations of HESS~J1800-240A/B/C (the yellow circles in Figure~\ref{hwc_lat}). 

To compare the $\gamma-$ray excesses as observed by {\it Fermi} and HAWC, we show a close-up view of the $3^{\circ}\times3^{\circ}$ region centered on UKS 1 as observed by LAT with the 3HWC significance contours overlaid in the lower panel of Figure~\ref{hwc_lat}. Both excesses are apparently extended in a similar orientation towards the Galactic plane (the red dashed line in the top panel of Figure~\ref{hwc_lat}). 

Based on 3HWC survey data, the differential flux at the peak position of the TeV feature is found to be $f^{\rm 3HWC}_{\gamma}=(3.27\pm1.61)\times10^{-14}$~cm$^{-2}$~s$^{-1}$~TeV$^{-1}$. Assuming the differential flux is at 7~TeV, we have also plotted the TeV flux together with the {\it Fermi}-LAT SED in Figure~\ref{sed} for comparison.  

\section{Summary and Discussion}

In this Letter, we report the detection of a GeV $\gamma$-ray feature at the location of the GC UKS 1 which may be associated with a TeV excess identified in the 3HWC survey data. 
While the TeV feature is well-resolved from other known VHE sources, it lies in a complex region of the $\gamma$-ray sky~\citep{HESS_GPS}, intersecting with the W28 supernova remnant field \citep{hess} and 
viewed through the Galactic bulge. As such, the possibility of source confusion or contamination cannot be firmly ruled out with current data, and follow-up VHE $\gamma-$ray observations are needed to improve confidence in the association between the observed GeV and TeV features.

We estimate the full-band GeV $\gamma-$ray luminosity of UKS 1 to be $L^{\rm LAT}_{\gamma}=(9.44\pm7.97)\times10^{35}$~erg~s$^{-1}$, of which 20 per cent is below 10 GeV and could be attributed to  
curvature radiation originating from its population of MSPs \cite[e.g.,][]{Venter2009ApJ}. Such a luminosity 
would imply the presence of a potentially large population of $119 \pm 100$ MSPs in UKS 1, if they have an average spin-down power of 
$\langle \dot{E} \rangle \sim 2\times10^{34}$ erg s$^{-1}$ 
and characteristic $\gamma$-ray conversion efficiency of $\sim 0.08$ \citep{GC_LAT}. 
This population may operate as a powerful engine for the production of relativistic leptons and driving collective pulsar winds. 
However, the high velocity of UKS 1 through the Galactic disk (approximately $\sim 270$ km s$^{-1}$; see \citealt{uks1_pm, Baumgardt2021MNRAS}) would place the resulting bow shock at $R_{\rm bs} \approx 0.6$ pc under typical Galactic disk conditions ($\sim 0.1$ cm$^{-3}$ for the warm ionized component of the interstellar medium), assuming efficient mixing of pulsar and stellar winds, mass-loaded at a rate of $\sim 10^{-6}$ M$_{\odot}$ yr$^{-1}$~\citep[][]{Bednarek_2014}. This distance is comparable to the core radius of UKS 1~\citep{uks_core_radius}, suggesting that the GC wind would be confined by the bow shock, forming a trailing magnetotail. This configuration would result in a distorted, non-spherical VHE emission morphology. 

Particles accelerated at this shock primarily cool via ICS in background interstellar/cosmological radiation and
optical radiation fields from GC stars. Given UKS 1’s relatively low stellar luminosity of $\sim 4 \times 10^4$ ${\rm L}_{\odot}$ (Galactic GCs usually range from $10^5 - 10^6$ ${\rm L}_{\odot}$; \citealt{Hilker2020IAUS}), the cooling length scale for particles at multi-TeV energies at the bow shock location is of order $\sim 100$ pc \citep[if adopting the relations in][]{Bednarek_2014}. This allows them to propagate deep into the GC’s magnetotail before losing significant energy. There, scattering in self-amplified magnetic turbulence isotropizes the particle pitch angle distribution so that their displaced VHE ICS emission becomes observable \citep[][]{Krumholz_2024}. 

This scenario is consistent with the observed 0.39$^{\circ}$ offset of the HAWC TeV emission peak, corresponding to a physical displacement of $\sim 100$ pc, if the emission is driven primarily by particles of energies in the range 1-10 TeV. It also accounts for the system’s relatively high TeV luminosity compared to the MSP spin-down power, as radiative ICS losses are not severe in the weak starlight of UKS 1. 
If the $\gtrsim 10$ GeV emission is dominated by ICS and the spectral shape resembles that of the VHE-confirmed Ter 5, we estimate the full-band TeV ICS luminosity of UKS 1 may be as high as $L_{\gamma}^{\rm ICS} \approx (9.02 \pm 4.43) \times 10^{35}$ erg s$^{-1}$. For comparison, the starlight intensity in the direction of Ter 5 is much stronger, leading to more radiative cooling and a lower TeV-to-GeV flux ratio. As a result, the detected TeV-to-GeV flux ratio of UKS 1 is an order of magnitude higher than that of Ter 5 \citep[see][]{Kong_2010, Ter5_HESS}.

This interpretation sets UKS 1 apart from the typical GCs in our Galaxy. In particular, its expected large MSP population and relatively low stellar luminosity drives unusually bright GeV emission, substantial lepton injection, and especially favorable conditions for extended, offset TeV emission. If follow-up observations confirm UKS 1 as the origin of the 3HWC $\gamma$-ray excess, it would not only firmly establish GCs as a distinct VHE source class, but also provide a natural laboratory to test cosmic ray transport theories with its particularly extended TeV emission.

\begin{acknowledgements}
 C.Y.H. is supported by the research fund of Chungnam National University and by the National Research Foundation of Korea grant 2022R1F1A1073952. E.R.O. is supported by the RIKEN Special Postdoctoral Researcher Program for junior scientists, and also acknowledges support from the Japan Society for the Promotion of Science (JSPS) as a JSPS International Research Fellow while at Osaka University (grant number JP22F22327). 
\end{acknowledgements}

\bibliographystyle{aa}
\bibliography{reference}

\begin{thebibliography}{30}
\expandafter\ifx\csname natexlab\endcsname\relax\def\natexlab#1{#1}\fi

\bibitem[{{Abdo} {et~al.}(2009{\natexlab{a}}){Abdo}, {Ackermann}, {Ajello},
  {Atwood}, {Axelsson}, {Baldini}, {Ballet}, {Barbiellini}, {Bastieri},
  {Baughman}, {Bechtol}, {Bellazzini}, {Berenji}, {Blandford}, {Bloom},
  {Bonamente}, {Borgland}, {Bregeon}, {Brez}, {Brigida}, {Bruel}, {Burnett},
  {Caliandro}, {Cameron}, {Caraveo}, {Casandjian}, {Cecchi}, {{\c{C}}elik},
  {Charles}, {Chaty}, {Chekhtman}, {Cheung}, {Chiang}, {Ciprini}, {Claus},
  {Cohen-Tanugi}, {Conrad}, {Cutini}, {Dermer}, {de Palma}, {Digel}, {Dormody},
  {do Couto e Silva}, {Drell}, {Dubois}, {Dumora}, {Farnier}, {Favuzzi},
  {Fegan}, {Focke}, {Frailis}, {Fukazawa}, {Fusco}, {Gargano}, {Gasparrini},
  {Gehrels}, {Germani}, {Giebels}, {Giglietto}, {Giordano}, {Glanzman},
  {Godfrey}, {Grenier}, {Grove}, {Guillemot}, {Guiriec}, {Hanabata}, {Harding},
  {Hayashida}, {Hays}, {Horan}, {Hughes}, {J{\'o}hannesson}, {Johnson},
  {Johnson}, {Johnson}, {Johnson}, {Kamae}, {Katagiri}, {Kawai}, {Kerr},
  {Kn{\"o}dlseder}, {Kuehn}, {Kuss}, {Lande}, {Latronico}, {Lemoine-Goumard},
  {Longo}, {Loparco}, {Lott}, {Lovellette}, {Lubrano}, {Makeev}, {Mazziotta},
  {McConville}, {McEnery}, {Meurer}, {Michelson}, {Mitthumsiri}, {Mizuno},
  {Moiseev}, {Monte}, {Monzani}, {Morselli}, {Moskalenko}, {Murgia}, {Nolan},
  {Norris}, {Nuss}, {Ohsugi}, {Omodei}, {Orlando}, {Ormes}, {Paneque},
  {Panetta}, {Parent}, {Pelassa}, {Pepe}, {Pierbattista}, {Piron}, {Porter},
  {Rain{\`o}}, {Rando}, {Razzano}, {Rea}, {Reimer}, {Reimer}, {Reposeur},
  {Ritz}, {Rochester}, {Rodriguez}, {Romani}, {Roth}, {Ryde}, {Sadrozinski},
  {Sanchez}, {Sander}, {Saz Parkinson}, {Sgr{\`o}}, {Smith}, {Smith},
  {Spandre}, {Spinelli}, {Starck}, {Strickman}, {Suson}, {Tajima}, {Takahashi},
  {Tanaka}, {Thayer}, {Thayer}, {Thompson}, {Tibaldo}, {Torres}, {Tosti},
  {Tramacere}, {Uchiyama}, {Usher}, {Vasileiou}, {Vilchez}, {Vitale}, {Wang},
  {Webb}, {Winer}, {Wood}, {Ylinen}, \& {Ziegler}}]{47Tuc_LAT}
{Abdo}, A.~A., {Ackermann}, M., {Ajello}, M., {et~al.} 2009{\natexlab{a}},
  Science, 325, 845

\bibitem[{{Abdo} {et~al.}(2009{\natexlab{b}}){Abdo}, {Ackermann}, {Ajello},
  {Atwood}, {Axelsson}, {Baldini}, {Ballet}, {Barbiellini}, {Bastieri},
  {Baughman}, {Bechtol}, {Bellazzini}, {Berenji}, {Bloom}, {Bonamente},
  {Borgland}, {Bregeon}, {Brez}, {Brigida}, {Bruel}, {Burnett}, {Caliandro},
  {Cameron}, {Caraveo}, {Carlson}, {Casandjian}, {Cecchi}, {{\c{C}}elik},
  {Chekhtman}, {Cheung}, {Ciprini}, {Claus}, {Cohen-Tanugi}, {Conrad},
  {Cutini}, {Dermer}, {de Angelis}, {de Palma}, {Digel}, {Silva}, {Drell},
  {Dubois}, {Dumora}, {Farnier}, {Favuzzi}, {Fegan}, {Focke}, {Frailis},
  {Fukazawa}, {Funk}, {Fusco}, {Gargano}, {Gasparrini}, {Gehrels}, {Germani},
  {Giebels}, {Giglietto}, {Giordano}, {Glanzman}, {Godfrey}, {Grenier},
  {Grondin}, {Grove}, {Guillemot}, {Guiriec}, {Hanabata}, {Harding},
  {Hayashida}, {Hays}, {Hughes}, {J{\'o}hannesson}, {Johnson}, {Johnson},
  {Johnson}, {Kamae}, {Katagiri}, {Kawai}, {Kerr}, {Kn{\"o}dlseder}, {Kocian},
  {Kuehn}, {Kuss}, {Lande}, {Latronico}, {Lemoine-Goumard}, {Longo}, {Loparco},
  {Lott}, {Lovellette}, {Lubrano}, {Makeev}, {Mazziotta}, {McEnery}, {Meurer},
  {Michelson}, {Mitthumsiri}, {Mizuno}, {Moiseev}, {Monte}, {Monzani},
  {Morselli}, {Moskalenko}, {Murgia}, {Nolan}, {Norris}, {Nuss}, {Ohsugi},
  {Okumura}, {Omodei}, {Orlando}, {Ormes}, {Ozaki}, {Paneque}, {Panetta},
  {Parent}, {Pepe}, {Pesce-Rollins}, {Piron}, {Pohl}, {Porter}, {Rain{\`o}},
  {Rando}, {Razzano}, {Reimer}, {Reimer}, {Reposeur}, {Ritz}, {Rochester},
  {Rodriguez}, {Ryde}, {Sadrozinski}, {Sanchez}, {Sander}, {Saz Parkinson},
  {Schalk}, {Sellerholm}, {Sgr{\`o}}, {Smith}, {Smith}, {Spandre}, {Spinelli},
  {Starck}, {Stecker}, {Strickman}, {Strong}, {Suson}, {Tajima}, {Takahashi},
  {Takahashi}, {Tanaka}, {Thayer}, {Thayer}, {Thompson}, {Tibaldo}, {Torres},
  {Tosti}, {Tramacere}, {Uchiyama}, {Usher}, {Vasileiou}, {Vilchez}, {Vitale},
  {Waite}, {Wang}, {Winer}, {Wood}, {Ylinen}, \& {Ziegler}}]{gamma_bkg}
{Abdo}, A.~A., {Ackermann}, M., {Ajello}, M., {et~al.} 2009{\natexlab{b}},
  \apj, 703, 1249

\bibitem[{{Abdo} {et~al.}(2010){Abdo}, {Ackermann}, {Ajello}, {Baldini},
  {Ballet}, {Barbiellini}, {Bastieri}, {Bellazzini}, {Blandford}, {Bloom},
  {Bonamente}, {Borgland}, {Bouvier}, {Brandt}, {Bregeon}, {Brigida}, {Bruel},
  {Buehler}, {Buson}, {Caliandro}, {Cameron}, {Caraveo}, {Carrigan},
  {Casandjian}, {Charles}, {Chaty}, {Chekhtman}, {Cheung}, {Chiang}, {Ciprini},
  {Claus}, {Cohen-Tanugi}, {Conrad}, {Decesar}, {Dermer}, {de Palma}, {Digel},
  {Silva}, {Drell}, {Dubois}, {Dumora}, {Favuzzi}, {Fortin}, {Frailis},
  {Fukazawa}, {Fusco}, {Gargano}, {Gasparrini}, {Gehrels}, {Germani},
  {Giglietto}, {Giordano}, {Glanzman}, {Godfrey}, {Grenier}, {Grondin},
  {Grove}, {Guillemot}, {Guiriec}, {Hadasch}, {Harding}, {Hays}, {Jean},
  {J{\'o}hannesson}, {Johnson}, {Johnson}, {Kamae}, {Katagiri}, {Kataoka},
  {Kerr}, {Kn{\"o}dlseder}, {Kuss}, {Lande}, {Latronico}, {Lee},
  {Lemoine-Goumard}, {Llena Garde}, {Longo}, {Loparco}, {Lovellette},
  {Lubrano}, {Makeev}, {Mazziotta}, {Michelson}, {Mitthumsiri}, {Mizuno},
  {Monte}, {Monzani}, {Morselli}, {Moskalenko}, {Murgia}, {Naumann-Godo},
  {Nolan}, {Norris}, {Nuss}, {Ohsugi}, {Omodei}, {Orlando}, {Ormes},
  {Pancrazi}, {Parent}, {Pepe}, {Pesce-Rollins}, {Piron}, {Porter},
  {Rain{\`o}}, {Rando}, {Reimer}, {Reimer}, {Reposeur}, {Ripken}, {Romani},
  {Roth}, {Sadrozinski}, {Saz Parkinson}, {Sgr{\`o}}, {Siskind}, {Smith},
  {Spinelli}, {Strickman}, {Suson}, {Takahashi}, {Takahashi}, {Tanaka},
  {Thayer}, {Thayer}, {Tibaldo}, {Torres}, {Tosti}, {Tramacere}, {Uchiyama},
  {Usher}, {Vasileiou}, {Venter}, {Vilchez}, {Vitale}, {Waite}, {Wang}, {Webb},
  {Winer}, {Yang}, {Ylinen}, {Ziegler}, \& {Fermi LAT Collaboration}}]{GC_LAT}
{Abdo}, A.~A., {Ackermann}, M., {Ajello}, M., {et~al.} 2010, \aap, 524, A75

\bibitem[{{Aharonian} {et~al.}(2008){Aharonian}, {Akhperjanian}, {Bazer-Bachi},
  {Behera}, {Beilicke}, {Benbow}, {Berge}, {Bernl{\"o}hr}, {Boisson}, {Bolz},
  {Borrel}, {Braun}, {Brion}, {Brown}, {B{\"u}hler}, {Bulik}, {B{\"u}sching},
  {Boutelier}, {Carrigan}, {Chadwick}, {Chounet}, {Clapson}, {Coignet},
  {Cornils}, {Costamante}, {Degrange}, {Dickinson}, {Djannati-Ata{\"\i}},
  {Domainko}, {O'C. Drury}, {Dubus}, {Dyks}, {Egberts}, {Emmanoulopoulos},
  {Espigat}, {Farnier}, {Feinstein}, {Fiasson}, {F{\"o}rster}, {Fontaine},
  {Fukui}, {Funk}, {Funk}, {F{\"u}{\ss}ling}, {Gallant}, {Giebels},
  {Glicenstein}, {Gl{\"u}ck}, {Goret}, {Hadjichristidis}, {Hauser}, {Hauser},
  {Heinzelmann}, {Henri}, {Hermann}, {Hinton}, {Hoffmann}, {Hofmann},
  {Holleran}, {Hoppe}, {Horns}, {Jacholkowska}, {de Jager}, {Kendziorra},
  {Kerschhaggl}, {Kh{\'e}lifi}, {Komin}, {Kosack}, {Lamanna}, {Latham}, {Le
  Gallou}, {Lemi{\`e}re}, {Lemoine-Goumard}, {Lenain}, {Lohse}, {Martin},
  {Martineau-Huynh}, {Marcowith}, {Masterson}, {Maurin}, {McComb}, {Moderski},
  {Moriguchi}, {Moulin}, {de Naurois}, {Nedbal}, {Nolan}, {Olive}, {Orford},
  {Osborne}, {Ostrowski}, {Panter}, {Pedaletti}, {Pelletier}, {Petrucci},
  {Pita}, {P{\"u}hlhofer}, {Punch}, {Ranchon}, {Raubenheimer}, {Raue},
  {Rayner}, {Reimer}, {Renaud}, {Ripken}, {Rob}, {Rolland}, {Rosier-Lees},
  {Rowell}, {Rudak}, {Ruppel}, {Sahakian}, {Santangelo}, {Saug{\'e}},
  {Schlenker}, {Schlickeiser}, {Schr{\"o}der}, {Schwanke}, {Schwarzburg},
  {Schwemmer}, {Shalchi}, {Sol}, {Spangler}, {Stawarz}, {Steenkamp},
  {Stegmann}, {Superina}, {Takeuchi}, {Tam}, {Tavernet}, {Terrier}, {van
  Eldik}, {Vasileiadis}, {Venter}, {Vialle}, {Vincent}, {Vivier}, {V{\"o}lk},
  {Volpe}, {Wagner}, \& {Ward}}]{hess}
{Aharonian}, F., {Akhperjanian}, A.~G., {Bazer-Bachi}, A.~R., {et~al.} 2008,
  \aap, 481, 401

\bibitem[{{Albert} {et~al.}(2020){Albert}, {Alfaro}, {Alvarez}, {Camacho},
  {Arteaga-Vel{\'a}zquez}, {Arunbabu}, {Avila Rojas}, {Ayala Solares},
  {Baghmanyan}, {Belmont-Moreno}, {BenZvi}, {Brisbois}, {Caballero-Mora},
  {Capistr{\'a}n}, {Carrami{\~n}ana}, {Casanova}, {Cotti}, {Couti{\~n}o de
  Le{\'o}n}, {De la Fuente}, {Diaz Hernandez}, {Diaz-Cruz}, {Dingus},
  {DuVernois}, {Durocher}, {D{\'\i}az-V{\'e}lez}, {Ellsworth}, {Engel},
  {Espinoza}, {Fan}, {Fang}, {Alonso}, {Fleischhack}, {Fraija},
  {Galv{\'a}n-G{\'a}mez}, {Garcia}, {Garc{\'\i}a-Gonz{\'a}lez}, {Garfias},
  {Giacinti}, {Gonz{\'a}lez}, {Goodman}, {Harding}, {Hernandez}, {Hinton},
  {Hona}, {Huang}, {Hueyotl-Zahuantitla}, {H{\"u}ntemeyer}, {Iriarte},
  {Jardin-Blicq}, {Joshi}, {Kieda}, {Lara}, {Lee}, {Le{\'o}n Vargas},
  {Linnemann}, {Longinotti}, {Luis-Raya}, {Lundeen}, {L{\'o}pez-Coto},
  {Malone}, {Marandon}, {Martinez}, {Martinez-Castellanos},
  {Mart{\'\i}nez-Castro}, {Matthews}, {Miranda-Romagnoli}, {Morales-Soto},
  {Moreno}, {Mostaf{\'a}}, {Nayerhoda}, {Nellen}, {Newbold}, {Nisa},
  {Noriega-Papaqui}, {Olivera-Nieto}, {Omodei}, {Peisker}, {P{\'e}rez Araujo},
  {P{\'e}rez-P{\'e}rez}, {Ren}, {Rho}, {Rivi{\`e}re}, {Rosa-Gonz{\'a}lez},
  {Ruiz-Velasco}, {Salazar}, {Salesa Greus}, {Sandoval}, {Schneider},
  {Schoorlemmer}, {Serna}, {Sinnis}, {Smith}, {Springer}, {Surajbali},
  {Tollefson}, {Torres}, {Torres-Escobedo}, {Ukwatta}, {Ure{\~n}a-Mena},
  {Weisgarber}, {Werner}, {Willox}, {Zepeda}, {Zhou}, {de Le{\'o}n},
  {{\'A}lvarez}, \& {HAWC Collaboration}}]{3hwc}
{Albert}, A., {Alfaro}, R., {Alvarez}, C., {et~al.} 2020, \apj, 905, 76

\bibitem[{{Ballet} {et~al.}(2023){Ballet}, {Bruel}, {Burnett}, {Lott}, \& {The
  Fermi-LAT collaboration}}]{4fgldr4}
{Ballet}, J., {Bruel}, P., {Burnett}, T.~H., {Lott}, B., \& {The Fermi-LAT
  collaboration}. 2023, arXiv e-prints, arXiv:2307.12546

\bibitem[{{Baumgardt} \& {Vasiliev}(2021)}]{Baumgardt2021MNRAS}
{Baumgardt}, H. \& {Vasiliev}, E. 2021, \mnras, 505, 5957

\bibitem[{{Bednarek} \& {Sobczak}(2014)}]{Bednarek_2014}
{Bednarek}, W. \& {Sobczak}, T. 2014, \mnras, 445, 2842

\bibitem[{{Cheng} {et~al.}(2010){Cheng}, {Chernyshov}, {Dogiel}, {Hui}, \&
  {Kong}}]{Cheng_2010}
{Cheng}, K.~S., {Chernyshov}, D.~O., {Dogiel}, V.~A., {Hui}, C.~Y., \& {Kong},
  A.~K.~H. 2010, \apj, 723, 1219

\bibitem[{{Fern{\'a}ndez-Trincado} {et~al.}(2020){Fern{\'a}ndez-Trincado},
  {Minniti}, {Beers}, {Villanova}, {Geisler}, {Souza}, {Smith}, {Placco},
  {Vieira}, {P{\'e}rez-Villegas}, {Barbuy}, {Alves-Brito}, {Bidin},
  {Alonso-Garc{\'\i}a}, {Tang}, \& {Palma}}]{uks1_pm}
{Fern{\'a}ndez-Trincado}, J.~G., {Minniti}, D., {Beers}, T.~C., {et~al.} 2020,
  \aap, 643, A145

\bibitem[{{Freire} {et~al.}(2011){Freire}, {Abdo}, {Ajello}, {Allafort},
  {Ballet}, {Barbiellini}, {Bastieri}, {Bechtol}, {Bellazzini}, {Blandford},
  {Bloom}, {Bonamente}, {Borgland}, {Brigida}, {Bruel}, {Buehler}, {Buson},
  {Caliandro}, {Cameron}, {Camilo}, {Caraveo}, {Cecchi}, {{\c{C}}elik},
  {Charles}, {Chekhtman}, {Cheung}, {Chiang}, {Ciprini}, {Claus}, {Cognard},
  {Cohen-Tanugi}, {Cominsky}, {de Palma}, {Dermer}, {do Couto e Silva},
  {Dormody}, {Drell}, {Dubois}, {Dumora}, {Espinoza}, {Favuzzi}, {Fegan},
  {Ferrara}, {Focke}, {Fortin}, {Fukazawa}, {Fusco}, {Gargano}, {Gasparrini},
  {Gehrels}, {Germani}, {Giglietto}, {Giordano}, {Giroletti}, {Glanzman},
  {Godfrey}, {Grenier}, {Grondin}, {Grove}, {Guillemot}, {Guiriec}, {Hadasch},
  {Harding}, {J{\'o}hannesson}, {Johnson}, {Johnson}, {Johnston}, {Katagiri},
  {Kataoka}, {Keith}, {Kerr}, {Kn{\"o}dlseder}, {Kramer}, {Kuss}, {Lande},
  {Latronico}, {Lee}, {Lemoine-Goumard}, {Longo}, {Loparco}, {Lovellette},
  {Lubrano}, {Lyne}, {Manchester}, {Marelli}, {Mazziotta}, {McEnery},
  {Michelson}, {Mizuno}, {Moiseev}, {Monte}, {Monzani}, {Morselli},
  {Moskalenko}, {Murgia}, {Nakamori}, {Nolan}, {Norris}, {Nuss}, {Ohsugi},
  {Okumura}, {Omodei}, {Orlando}, {Ozaki}, {Paneque}, {Parent},
  {Pesce-Rollins}, {Pierbattista}, {Piron}, {Porter}, {Rain{\`o}}, {Ransom},
  {Ray}, {Reimer}, {Reimer}, {Reposeur}, {Ritz}, {Romani}, {Roth},
  {Sadrozinski}, {Saz Parkinson}, {Shannon}, {Siskind}, {Smith}, {Spinelli},
  {Stappers}, {Suson}, {Takahashi}, {Tanaka}, {Tauris}, {Thayer}, {Theureau},
  {Thompson}, {Thorsett}, {Tibaldo}, {Torres}, {Tosti}, {Troja},
  {Vandenbroucke}, {Van Etten}, {Vasileiou}, {Venter}, {Vianello}, {Vilchez},
  {Vitale}, {Waite}, {Wang}, {Wood}, {Yang}, {Ziegler}, \& {Zimmer}}]{NGC6624A}
{Freire}, P.~C.~C., {Abdo}, A.~A., {Ajello}, M., {et~al.} 2011, Science, 334,
  1107

\bibitem[{{H.~E.~S.~S. Collaboration} {et~al.}(2018){H.~E.~S.~S.
  Collaboration}, {Abdalla}, {Abramowski}, {Aharonian}, {Ait Benkhali},
  {Ang{\"u}ner}, {Arakawa}, {Arrieta}, {Aubert}, {Backes}, {Balzer}, {Barnard},
  {Becherini}, {Becker Tjus}, {Berge}, {Bernhard}, {Bernl{\"o}hr}, {Blackwell},
  {B{\"o}ttcher}, {Boisson}, {Bolmont}, {Bonnefoy}, {Bordas}, {Bregeon},
  {Brun}, {Brun}, {Bryan}, {B{\"u}chele}, {Bulik}, {Capasso}, {Carrigan},
  {Caroff}, {Carosi}, {Casanova}, {Cerruti}, {Chakraborty}, {Chaves}, {Chen},
  {Chevalier}, {Colafrancesco}, {Condon}, {Conrad}, {Davids}, {Decock}, {Deil},
  {Devin}, {deWilt}, {Dirson}, {Djannati-Ata{\"\i}}, {Domainko}, {Donath},
  {Drury}, {Dutson}, {Dyks}, {Edwards}, {Egberts}, {Eger}, {Emery},
  {Ernenwein}, {Eschbach}, {Farnier}, {Fegan}, {Fernandes}, {Fiasson},
  {Fontaine}, {F{\"o}rster}, {Funk}, {F{\"u}{\ss}ling}, {Gabici}, {Gallant},
  {Garrigoux}, {Gast}, {Gat{\'e}}, {Giavitto}, {Giebels}, {Glawion},
  {Glicenstein}, {Gottschall}, {Grondin}, {Hahn}, {Haupt}, {Hawkes},
  {Heinzelmann}, {Henri}, {Hermann}, {Hinton}, {Hofmann}, {Hoischen}, {Holch},
  {Holler}, {Horns}, {Ivascenko}, {Iwasaki}, {Jacholkowska}, {Jamrozy},
  {Jankowsky}, {Jankowsky}, {Jingo}, {Jouvin}, {Jung-Richardt}, {Kastendieck},
  {Katarzy{\'n}ski}, {Katsuragawa}, {Katz}, {Kerszberg}, {Khangulyan},
  {Kh{\'e}lifi}, {King}, {Klepser}, {Klochkov}, {Klu{\'z}niak}, {Komin},
  {Kosack}, {Krakau}, {Kraus}, {Kr{\"u}ger}, {Laffon}, {Lamanna}, {Lau},
  {Lees}, {Lefaucheur}, {Lemi{\`e}re}, {Lemoine-Goumard}, {Lenain}, {Leser},
  {Lohse}, {Lorentz}, {Liu}, {L{\'o}pez-Coto}, {Lypova}, {Marandon},
  {Malyshev}, {Marcowith}, {Mariaud}, {Marx}, {Maurin}, {Maxted}, {Mayer},
  {Meintjes}, {Meyer}, {Mitchell}, {Moderski}, {Mohamed}, {Mohrmann},
  {Mor{\r{a}}}, {Moulin}, {Murach}, {Nakashima}, {de Naurois}, {Ndiyavala},
  {Niederwanger}, {Niemiec}, {Oakes}, {O'Brien}, {Odaka}, {Ohm}, {Ostrowski},
  {Oya}, {Padovani}, {Panter}, {Parsons}, {Paz Arribas}, {Pekeur}, {Pelletier},
  {Perennes}, {Petrucci}, {Peyaud}, {Piel}, {Pita}, {Poireau}, {Poon},
  {Prokhorov}, {Prokoph}, {P{\"u}hlhofer}, {Punch}, {Quirrenbach}, {Raab},
  {Rauth}, {Reimer}, {Reimer}, {Renaud}, {de los Reyes}, {Rieger}, {Rinchiuso},
  {Romoli}, {Rowell}, {Rudak}, {Rulten}, {Safi-Harb}, {Sahakian}, {Saito},
  {Sanchez}, {Santangelo}, {Sasaki}, {Schandri}, {Schlickeiser},
  {Sch{\"u}ssler}, {Schulz}, {Schwanke}, \& {Schwemmer}}]{HESS_GPS}
{H.~E.~S.~S. Collaboration}, {Abdalla}, H., {Abramowski}, A., {et~al.} 2018,
  \aap, 612, A1

\bibitem[{{H.~E.~S.~S. Collaboration} {et~al.}(2011){H.~E.~S.~S.
  Collaboration}, {Abramowski}, {Acero}, {Aharonian}, {Akhperjanian}, {Anton},
  {Balzer}, {Barnacka}, {Barres de Almeida}, {Becherini}, {Becker}, {Behera},
  {Bernl{\"o}hr}, {Bochow}, {Boisson}, {Bolmont}, {Bordas}, {Brucker}, {Brun},
  {Brun}, {Bulik}, {B{\"u}sching}, {Carrigan}, {Casanova}, {Cerruti},
  {Chadwick}, {Charbonnier}, {Chaves}, {Cheesebrough}, {Chounet}, {Clapson},
  {Coignet}, {Cologna}, {Conrad}, {Dalton}, {Daniel}, {Davids}, {Degrange},
  {Deil}, {Dickinson}, {Djannati-Ata{\"\i}}, {Domainko}, {Drury}, {Dubois},
  {Dubus}, {Dutson}, {Dyks}, {Dyrda}, {Egberts}, {Eger}, {Espigat}, {Fallon},
  {Farnier}, {Fegan}, {Feinstein}, {Fernandes}, {Fiasson}, {Fontaine},
  {F{\"o}rster}, {F{\"u}{\ss}ling}, {Gallant}, {Gast}, {G{\'e}rard}, {Gerbig},
  {Giebels}, {Glicenstein}, {Gl{\"u}ck}, {Goret}, {G{\"o}ring}, {H{\"a}ffner},
  {Hague}, {Hampf}, {Hauser}, {Heinz}, {Heinzelmann}, {Henri}, {Hermann},
  {Hinton}, {Hoffmann}, {Hofmann}, {Hofverberg}, {Holler}, {Horns},
  {Jacholkowska}, {de Jager}, {Jahn}, {Jamrozy}, {Jung}, {Kastendieck},
  {Katarzy{\'n}ski}, {Katz}, {Kaufmann}, {Keogh}, {Khangulyan}, {Kh{\'e}lifi},
  {Klochkov}, {Klu{\'z}niak}, {Kneiske}, {Komin}, {Kosack}, {Kossakowski},
  {Laffon}, {Lamanna}, {Lennarz}, {Lohse}, {Lopatin}, {Lu}, {Marandon},
  {Marcowith}, {Masbou}, {Maurin}, {Maxted}, {McComb}, {Medina}, {M{\'e}hault},
  {Moderski}, {Moulin}, {Naumann}, {Naumann-Godo}, {de Naurois}, {Nedbal},
  {Nekrassov}, {Nguyen}, {Nicholas}, {Niemiec}, {Nolan}, {Ohm}, {de O{\~n}a
  Wilhelmi}, {Opitz}, {Ostrowski}, {Oya}, {Panter}, {Paz Arribas}, {Pedaletti},
  {Pelletier}, {Petrucci}, {Pita}, {P{\"u}hlhofer}, {Punch}, {Quirrenbach},
  {Raue}, {Rayner}, {Reimer}, {Reimer}, {Renaud}, {de los Reyes}, {Rieger},
  {Ripken}, {Rob}, {Rosier-Lees}, {Rowell}, {Rudak}, {Rulten}, {Ruppel},
  {Ryde}, {Sahakian}, {Santangelo}, {Schlickeiser}, {Sch{\"o}ck}, {Schulz},
  {Schwanke}, {Schwarzburg}, {Schwemmer}, {Sikora}, {Skilton}, {Sol},
  {Spengler}, {Stawarz}, {Steenkamp}, {Stegmann}, {Stinzing}, {Stycz},
  {Sushch}, {Szostek}, {Tavernet}, {Terrier}, {Tluczykont}, {Valerius}, {van
  Eldik}, {Vasileiadis}, {Venter}, {Vialle}, {Viana}, {Vincent}, {V{\"o}lk},
  {Volpe}, {Vorobiov}, {Vorster}, {Wagner}, {Ward}, {White}, {Wierzcholska},
  {Zacharias}, {Zajczyk}, {Zdziarski}, {Zech}, \& {Zechlin}}]{Ter5_HESS}
{H.~E.~S.~S. Collaboration}, {Abramowski}, A., {Acero}, F., {et~al.} 2011,
  \aap, 531, L18

\bibitem[{{Harris}(1996)}]{Harris1996.112}
{Harris}, W.~E. 1996, \aj, 112, 1487

\bibitem[{{Henry} {et~al.}(2024){Henry}, {Paglione}, {Song}, {Tan}, {Zurek}, \&
  {Pinto}}]{2024stackGC}
{Henry}, O.~K., {Paglione}, T. A.~D., {Song}, Y., {et~al.} 2024, \mnras, 535,
  434

\bibitem[{{Hilker} {et~al.}(2020){Hilker}, {Baumgardt}, {Sollima}, \&
  {Bellini}}]{Hilker2020IAUS}
{Hilker}, M., {Baumgardt}, H., {Sollima}, A., \& {Bellini}, A. 2020, in IAU
  Symposium, Vol. 351, Star Clusters: From the Milky Way to the Early Universe,
  ed. A.~{Bragaglia}, M.~{Davies}, A.~{Sills}, \& E.~{Vesperini}, 451--454

\bibitem[{{Hui} {et~al.}(2010){Hui}, {Cheng}, \& {Taam}}]{Hui2010.714}
{Hui}, C.~Y., {Cheng}, K.~S., \& {Taam}, R.~E. 2010, \apj, 714, 1149

\bibitem[{{Hui} {et~al.}(2011){Hui}, {Cheng}, {Wang}, {Tam}, {Kong},
  {Chernyshov}, \& {Dogiel}}]{Hui_2011}
{Hui}, C.~Y., {Cheng}, K.~S., {Wang}, Y., {et~al.} 2011, \apj, 726, 100

\bibitem[{{Kong} {et~al.}(2010){Kong}, {Hui}, \& {Cheng}}]{Kong_2010}
{Kong}, A.~K.~H., {Hui}, C.~Y., \& {Cheng}, K.~S. 2010, \apjl, 712, L36

\bibitem[{Krumholz {et~al.}(2024)Krumholz, Crocker, Bahramian, \&
  Bordas}]{Krumholz_2024}
Krumholz, M., Crocker, R., Bahramian, A., \& Bordas, P. 2024, Nature Astronomy,
  8, 1284

\bibitem[{{Mattox} {et~al.}(1996){Mattox}, {Bertsch}, {Chiang}, {Dingus},
  {Digel}, {Esposito}, {Fierro}, {Hartman}, {Hunter}, {Kanbach}, {Kniffen},
  {Lin}, {Macomb}, {Mayer-Hasselwander}, {Michelson}, {von Montigny},
  {Mukherjee}, {Nolan}, {Ramanamurthy}, {Schneid}, {Sreekumar}, {Thompson}, \&
  {Willis}}]{1996ApJ...461..396M}
{Mattox}, J.~R., {Bertsch}, D.~L., {Chiang}, J., {et~al.} 1996, \apj, 461, 396

\bibitem[{{McCarver} {et~al.}(2024){McCarver}, {Maccarone}, {Ransom}, {Clarke},
  {Giacintucci}, {Peters}, {Polisensky}, {Nyland}, {Gautam}, {Freire}, \&
  {Rangelov}}]{uks_core_radius}
{McCarver}, A.~V., {Maccarone}, T.~J., {Ransom}, S.~M., {et~al.} 2024, \apj,
  969, 30

\bibitem[{{Phan} {et~al.}(2020){Phan}, {Gabici}, {Morlino}, {Terrier}, {Vink},
  {Krause}, \& {Menu}}]{w28}
{Phan}, V.~H.~M., {Gabici}, S., {Morlino}, G., {et~al.} 2020, \aap, 635, A40

\bibitem[{{Song} {et~al.}(2021){Song}, {Macias}, {Horiuchi}, {Crocker}, \&
  {Nataf}}]{Song_2021}
{Song}, D., {Macias}, O., {Horiuchi}, S., {Crocker}, R.~M., \& {Nataf}, D.~M.
  2021, \mnras, 507, 5161

\bibitem[{{Tam} {et~al.}(2011){Tam}, {Kong}, {Hui}, {Cheng}, {Li}, \&
  {Lu}}]{Tam_2011}
{Tam}, P.~H.~T., {Kong}, A.~K.~H., {Hui}, C.~Y., {et~al.} 2011, \apj, 729, 90

\bibitem[{{Venter} {et~al.}(2009){Venter}, {De Jager}, \&
  {Clapson}}]{Venter2009ApJ}
{Venter}, C., {De Jager}, O.~C., \& {Clapson}, A.~C. 2009, \apjl, 696, L52

\bibitem[{{Wu} {et~al.}(2013){Wu}, {Hui}, {Wu}, {Kong}, {Huang}, {Tam},
  {Takata}, \& {Cheng}}]{Wu_2013}
{Wu}, J.~H.~K., {Hui}, C.~Y., {Wu}, E.~M.~H., {et~al.} 2013, \apjl, 765, L47

\bibitem[{Yuan {et~al.}(2022)Yuan, Zheng, Zhang, \& Zhang}]{Yuan_2022}
Yuan, M., Zheng, J., Zhang, P., \& Zhang, L. 2022, Research in Astronomy and
  Astrophysics, 22, 055019

\bibitem[{{Zhang} {et~al.}(2022){Zhang}, {Xing}, \& {Wang}}]{NGC6652B}
{Zhang}, P., {Xing}, Y., \& {Wang}, Z. 2022, \apjl, 935, L36

\bibitem[{{Zhang} {et~al.}(2023){Zhang}, {Xing}, {Wang}, {Wu}, \&
  {Chen}}]{M92A}
{Zhang}, P., {Xing}, Y., {Wang}, Z., {Wu}, W., \& {Chen}, Z. 2023, \apj, 945,
  70

\end{thebibliography}

\begin{appendix}
\renewcommand{\arraystretch}{1.1}
\section{Excesses around UKS 1 as observed by {\it Fermi} LAT\label{excess}}
\begin{figure}[h!]
    \hspace{-1.0cm}
    \includegraphics[width=11cm]{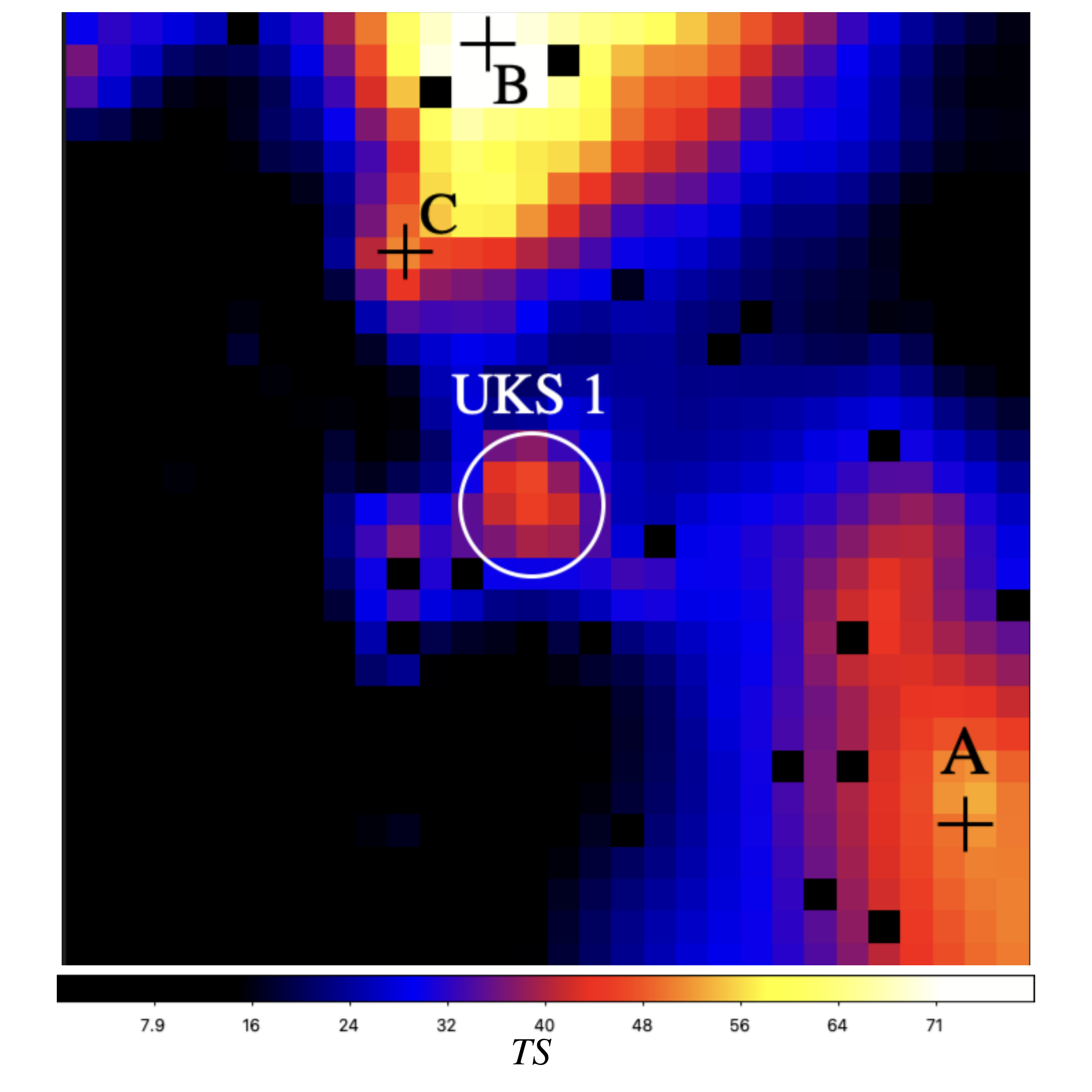}
    \vspace{-0.5cm}
    \caption{$3^{\circ}\times3^{\circ}$ {\it Fermi}-LAT TS map showing the $\gamma-$ray excess found around UKS~1.}
    \label{fig:excess}
\end{figure}

After subtracting contributions from the 4FGL sources, Galactic diffuse background and the extragalactic isotropic background, $\gamma-$ray excesses have been found in the field around UKS 1 (\autoref{fig:excess}). For better estimation of the detection significance and properties of UKS 1, these excesses have to be removed. 

To account for their contribution, we first re-ran the likelihood analysis, adding point sources at locations A and B (see \autoref{fig:excess}) to the model, each modeled with a power-law spectrum. However, an excess remained at location C. To address this, we added an additional point source at location C and repeated the analysis. The results are summarized in \autoref{tab:excess}.  
\FloatBarrier
\FloatBarrier
\begin{table}[h!]
    \centering
    \begin{tabular}{c|cccc}
    \hline
       Source  & RA (J2000) & Dec (J2000) & $\Gamma$ & $F_{\gamma}$ \\
               & degree & degree &  &  \\
       \hline
       A  & 267.172$^{\circ}$ & -25.172$^{\circ}$ &  $2.36\pm0.15$ & $2.32\pm2.82$ \\
        B  & 268.812$^{\circ}$ & -22.746$^{\circ}$ & $2.99\pm0.02$ & $7.49\pm0.30$\\
         C  & 269.092$^{\circ}$ &  -23.395$^{\circ}$ & $2.12\pm0.02$ & $1.79\pm0.30$ \\
    \hline
    \end{tabular}
    \caption{Locations and spectral parameters of additional point-source models used to account for and subtract the $\gamma$-ray excesses around UKS~1. $\Gamma$ is the photon index. Photon flux in 0.3-100 GeV $F_{\gamma}$ in unit of $10^{-9}$~photon~cm$^{-2}$~s$^{-1}$}
    \label{tab:excess}
\end{table}

\end{appendix}

\end{document}